\documentclass[%
 aip,
 amsmath,amssymb,
reprint,%
]{revtex4-1}

\usepackage{graphicx}
\usepackage{dcolumn}
\usepackage{bm}

\usepackage[utf8]{inputenc}
\usepackage[T1]{fontenc}
\usepackage{mathptmx}
\usepackage{etoolbox}
\usepackage{soul}
\usepackage{xcolor}

\begin{document}


\title{Measuring arrangement and size distributions of flowing droplets in microchannels using DropTrack}

\author{Mihir Durve}
\affiliation{Center for Life Nano- \& Neuro-Science, Fondazione Istituto Italiano di Tecnologia (IIT), viale Regina Elena 295, 00161 Rome, Italy}

\author{Sibilla Orsini}
\affiliation{NEST, Istituto Nanoscienze-CNR and Scuola Normale Superiore, Piazza San Silvestro 12, Pisa, 56127,  Italy}
\affiliation{Istituto per le Applicazioni del Calcolo del Consiglio Nazionale delle Ricerche, via dei Taurini 19, Roma, 00185, Italy}

\author{Adriano Tiribocchi}
\affiliation{Istituto per le Applicazioni del Calcolo del Consiglio Nazionale delle Ricerche, via dei Taurini 19, Roma, 00185, Italy}

\author{Andrea Montessori}
\affiliation{Dipartimento di Ingegneria, Università degli Studi Roma tre, via Vito Volterra 62, Rome, 00146, Italy}

\author{Jean-Michel Tucny}
\affiliation{Center for Life Nano- \& Neuro-Science, Fondazione Istituto Italiano di Tecnologia (IIT), viale Regina Elena 295, 00161 Rome, Italy}
\affiliation{Dipartimento di Ingegneria, Università degli Studi Roma tre, via Vito Volterra 62, Rome, 00146, Italy}

\author{Marco Lauricella$^{*}$}
\affiliation{Istituto per le Applicazioni del Calcolo del Consiglio Nazionale delle Ricerche, via dei Taurini 19, Roma, 00185, Italy}
\email{m.lauricella@cnr.it}

\author{Andrea Camposeo}
\affiliation{NEST, Istituto Nanoscienze-CNR and Scuola Normale Superiore, Piazza San Silvestro 12, Pisa, 56127,  Italy}

\author{Dario Pisignano}
\affiliation{NEST, Istituto Nanoscienze-CNR and Scuola Normale Superiore, Piazza San Silvestro 12, Pisa, 56127,  Italy}
\affiliation{ Dipartimento di Fisica, Università di Pisa, Largo B. Pontecorvo 3, Pisa, 56127, Italy}

\author{Sauro Succi}
\affiliation{Center for Life Nano- \& Neuro-Science, Fondazione Istituto Italiano di Tecnologia (IIT), viale Regina Elena 295, 00161 Rome, Italy}
\affiliation{Department of Physics, Harvard University, 17 Oxford St, Cambridge, MA 02138, United States}


\begin{abstract}
In microfluidic systems, droplets undergo intricate deformations as they traverse flow-focusing junctions, posing a challenging task for accurate measurement, especially during short transit times. This study investigates the physical behavior of droplets within dense emulsions in diverse microchannel geometries, specifically focusing on the impact of varying opening angles within the primary channel and injection rates of fluid components. Employing a sophisticated droplet tracking tool based on deep-learning techniques, we analyze multiple frames from flow-focusing experiments to quantitatively characterize droplet deformation in terms of ratio between maximum width and height and propensity to form liquid with hexagonal spatial arrangement. Our findings reveal the existence of an optimal opening angle where shape deformations are minimal and hexagonal arrangement is maximal.
Variations of fluid injection rates are also found to affect size and packing fraction of the emulsion in the exit channel. This paper offers insights into deformations, size and structure of fluid emulsions relative to microchannel geometry and other flow-related parameters captured through machine learning, with potential implications for the design of microchips utilized in cellular transport and tissue engineering applications.
\end{abstract}

\maketitle

\section{Introduction}

The high-throughput generation of fluid droplets holds significant importance in various fields, such as chemistry, biology and material science \cite{teh,hermingaus,abate}, and has found applications in different sectors of modern industries, ranging from food processing \cite{musch,skurt, he2020} and pharmaceutics for drug delivery \cite{xu,vladi,he2019} to nanoparticle synthesis \cite{dendukuri,walther} and soft bio-materials \cite{visser, Zhang2022}. 

Within the realm of microfluidics, the process can be realized through several techniques, which 
can be categorized into three primary types based on channel configurations: cross-flow, co-flow, and flow-focusing \cite{anna,teh,hermingaus}. These droplet-producing methods determine the droplet characteristics, such as their shape, size, surface properties,  
interactions, etc. \cite{Garstecki_1,milan2020lattice,sbragaglia,Pelusi_2019, tang2016} 
The co-flow and flow-focusing methods, in particular, are specifically adept at producing highly monodisperse droplets. These tiny elements are well suited for the assembly of droplet-based soft materials, such as foams \cite{marmottant2}, and are of interest in lab-on-chip devices \cite{dittrich}. In this respect, the precise control of droplet formation with predefined volume and size represents a formidable challenge since it hinges upon careful control of a number of factors including channel design and manufacturing, characteristics of the fluids (such as concentration, viscosity and surface tension), fine-tuning of flow rates as well as fluid-structure interaction \cite{anna,koster,diotallevi,montessori2,montessori3,kernakov,montessori4}. Despite the complexity of the process, it has been shown in a number of papers that the confinement of droplets brings out a variety of self-assembled ordered structures, ranging from foams to crystal-like templates with different degree of order, that would not exist without boundaries \cite{marmottant,marmottant2,garstecki}. 

Control over droplet sizes, deformations and their organization are essential for many applications. For example, in droplet-based drug delivery applications, it is crucial to maintain a sustainable production rate while ensuring a precise droplet volume. The size of the droplet plays a critical role in determining the drug release profile, as highlighted in multiple studies \cite{riahi2015, he2019}. Moreover, the manufacturing process can subject the delicate payload, such as probiotics, to significant environmental stresses which potentially can impact the stability and functionality of the drug delivery method \cite{Gurram2021}.

Considerable attention is given to the regulation of droplet size generation through the manipulation of microchannel geometry, fluid characteristics, and fluid injection rates \cite{abate2009,costa2017,Saqib2018,Sartipzadeh}. Careful examinations of droplet deformations are conducted within confined passages\cite{Saffar2023}. These studies have a predominant emphasis on individual droplets.  While many other studies have been focused on characterizing the dynamics and arrangement of fluid droplets in symmetric microfluidic channels in which the striction connects the exit channel by means of fixed opening angles \cite{weitz1,weitz2,vecchiolla}, much less is known about the influence of microchannel opening angle $\alpha$ on droplet size, their spatial disposition, and the resultant stresses endured by the droplets due to interactions with channel walls and neighboring droplets before reaching equilibrium in densely packed emulsions.

This work investigates the influence of microchannel geometry on the size and spatial arrangement of droplets within a flow-focusing apparatus. Advances in manufacturing methodologies \cite{su2023} for microfluidic devices, coupled with innovative computer vision algorithms, have facilitated the analysis of exceptionally high-quality data, allowing for the precise quantification of droplet characteristics across diverse configurations. Further, the fast analysis capabilities of computer vision algorithms reveal detailed information in the form of droplet size distributions rather than average values with other traditional measuring methods. Leveraging DropTrack \cite{Durve1,Durve2,Durve3,Durve4}, a previously established deep leaning-based droplet tracking tool, we quantify parameters related to droplet deformations, size distribution, packing density and arrangement under the confined environment of a microfluidic channel. 
We find that the geometry of the microchannel combined with the complex structure of the fluid flows critically affect the droplet-droplet interactions,  yielding an intricate picture where the opening angle of the channel and injection rates of the fluids control size and arrangement in a highly non-trivial manner.  Indeed, keeping fixed the rates of oil (continuous phase) and water (continuous dispersed phase), both deformation and average size of the droplets lessen for increasing values of opening angle. The latter is also found to affect droplet order, which exhibits an optimal hexagonal arrangement for a restricted range of opening angles. 
However, increasing the oil injection rate generally disrupts such order producing weakly-packed small droplets, essentially because of the high shear flows in the channel.  
Modification of the water injection rate has a milder impact on drop size, whereas it can considerably alter the resulting structure of the emulsion downstream.
The complex scenario emerging from these findings suggests that predicting the structural and mechanical properties of a fluid emulsion in a highly confined regime remains a challenge if solely based on the control of a limited set of physical parameters. Machine learning (ML) methods (potentially in combination with computer simulations) can effectively bridge this gap, especially if tested on different lab-on-chip platforms to minimize the dependence on a specific training dataset. 

The paper is structured as follows. In the next section, we illustrate the details of the experiments and the DropTrack algorithm. Afterwards, we present the results which contain a discussion about droplet deformability, size distribution, degree of order and packing fraction in channels with varying opening angles and oil/water injection rates. Some final considerations and potential perspectives close the manuscript.

\section{Experimental Details  and tracking algorithm structure}
{\label{exp_details}}

\subsection{Materials}
The materials used in the experiment are the following. The E-Shell®600 (EnvisionTEC) photocurable material was used as the pre-polymer. Also, Brilliant Black BN pigment was purchased from ABCR GmbH, Tween®20 and isopropyl alcohol were obtained from Sigma-Aldrich, while sunflower seed oil from Santa Cruz Biotechnology.

\subsection{Device for droplets generation}
\begin{figure*}
\begin{center}
\includegraphics[width=\linewidth, keepaspectratio]{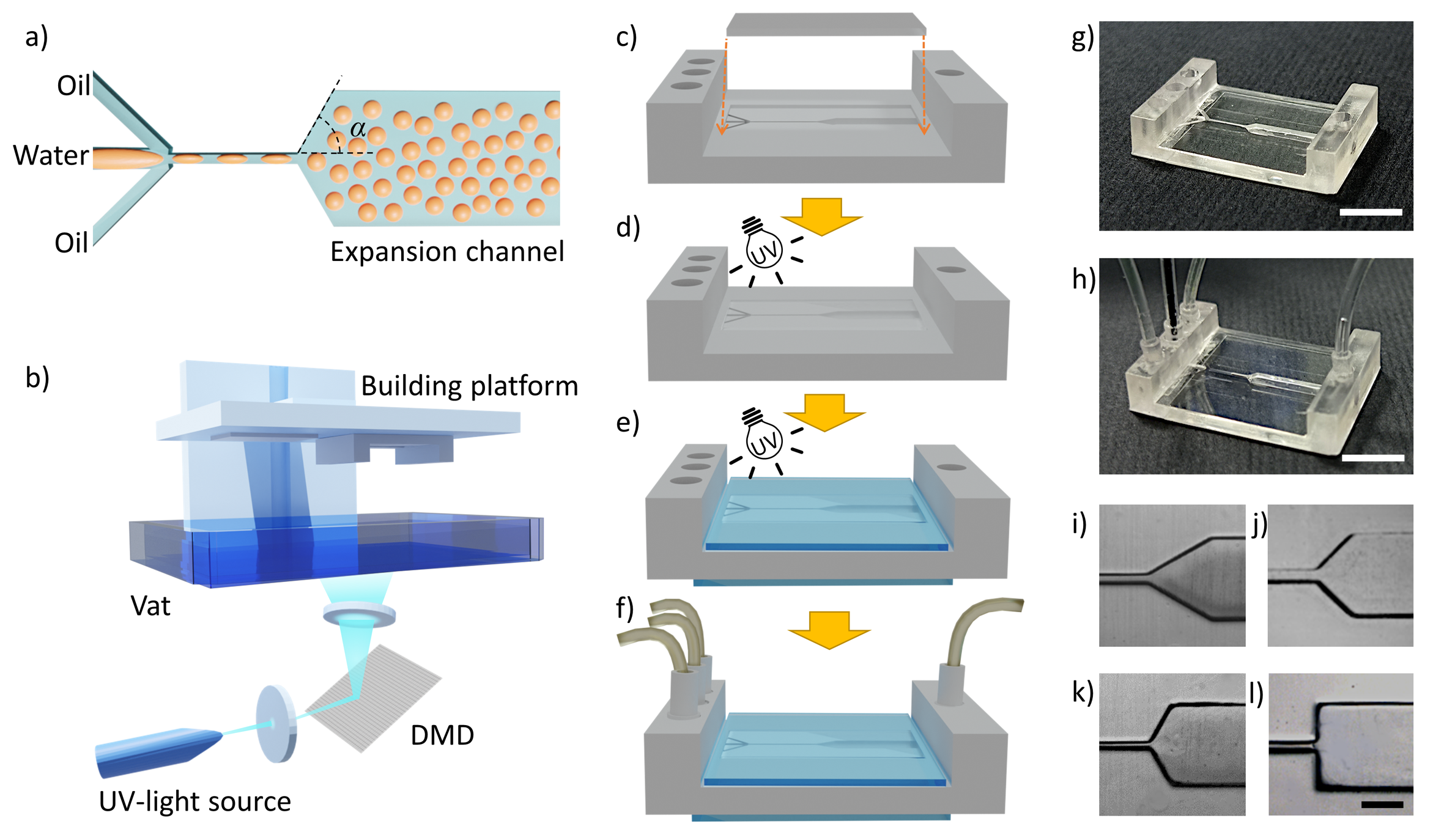}
\caption{a) Schematic representation of the flow-focusing junction and the expansion channel designed for the generation of W/O emulsion. b) Sketch of the DLP 3D printing system; DMD: Digital Micromirror device. c)-f) Illustration of the fabrication steps of the microfluidic device. First, the device with open channels and the sealing layer are printed (c) and bonded through UV-light irradiation. (d) Afterwards, two glass coverslips are bonded on the top and bottom surfaces of the device by UV light irradiation and photopolymerization of an interposed layer of pre-polymer (e). Finally, plastic tubes are connected to the inlets and outlet apertures by means of 3D printed connectors (f). g-h) Photographs of the device after step d), (g), and of the final device (h). Scale bars: 9 mm. i-l) Bright field optical microscope images of the inlet of the expansion chamber with opening angle: 30 (i), 45 (j), 60 (k) and 90 (l) degrees. Scale bar: 1 mm. \label{fig_device}}
\end{center}
\end{figure*}

A flow-focusing microfluidic junction was designed to produce water-in-oil (W/O) emulsion droplets (Figure \ref{fig_device}a). The device has a base sizes of 23.0x17.5 mm and a minimum thickness of 1.8 mm. The devices have one central inlet for the dispersed phase (water solution) and two side inlets for the continuous phase (sunflower oil). The latter channels form a 45° angle with respect to the central one. The inlet channels have a width of 500 $\mu$m. The flow-focusing junction is connected to an expansion channel with an opening angle ($\alpha$) of 30, 45, 60 or 90 degrees, respectively. The narrow channel has a width of 200 $\mu$m and a length of 5 mm, while the expansion channel has a width of 2 mm and a length of 9 mm. All channels have a depth of 500 $\mu$m. To prevent the presence in the channel of residual pre-polymer polymerized during the printing process, the channels were left open in the central area of the device (i.e. they were designed with one edge positioned on the surface of the device). In order to seal them, a 400 $\mu$m thick layer was designed as a cap for the channels. Either the inlet or outlet channels were connected to four cylindrical apertures with internal 2.2 mm diameter and 3.8 mm height, which host conical shaped hollow connectors used for inserting plastic tubes.

\subsection{Device fabrication}\label{dev_fab}

The device, the hollow connectors and the sealing layer were printed by a digital light processing (DLP) system (MicroPlus, HD EnvisionTEC) equipped with an UV light source with emission wavelength of 405 nm and intensity of 2 mW cm$^{-2}$ (Figure \ref{fig_device}b). The objects to be printed were sliced in 50 $\mu$m thick layers. A UV light exposure time of 4.8 s was used for each layer.
After the printing process, the device with open channels and the sealing layer were bonded through exposure to the emission of a UV light lamp (wavelength: 365 nm, intensity: 0.7 mW cm$^{-2}$, exposure time: 20 s), as illustrated in Figure \ref{fig_device}c-d. Afterwards, two glass coverslips (Bio-Optica, 22$\times$22 mm$^{2}$) were bonded to the top and bottom surface of the device, through the photopolymerization of a layer of E-Shell®600 interposed between the glass coverslips and the device surfaces (Figure \ref{fig_device}e). To this aim, the UV lamp (365 nm) was used (exposure time: 20 s). The resulting devices were washed with isopropyl alcohol and dried in a nitrogen flow. Finally, the hollow connectors and flexible plastic tubes were connected to the device (Figure \ref{fig_device}f).
A example of the realized device is shown in Figure \ref{fig_device}g-h, whereas optical microscope images of the inlet of the expansion channel with various opening angles are shown in Figure \ref{fig_device}i-l, respectively.

\subsection{Droplet formation}\label{drop_gen}
For obtaining W/O emulsions, Tween®20 (1 mg/ml) and Brilliant Black (7 mg/mL) were dissolved in deionized water as the surfactant and the black pigment, respectively. In particular, the latter was exploited to better visualize the emulsion formation and the droplets packing in the expansion channel. The flows of the continuous and dispersed phases through the device were controlled by a high-precision twin syringe pump (Harvard Apparatus). The droplets were imaged by using a stereo microscope (MZ 16 FA, Leica, illumination in transmission mode) and a camera (Fastcam APX RS, Photron, 3000 frames per second). The sequences of images were stored as AVI video files. Three different sequences of images were acquired for each set of experimental data.

\subsection{DropTrack algorithm}
\begin{figure}
\begin{center}
\includegraphics[width=\linewidth, keepaspectratio]{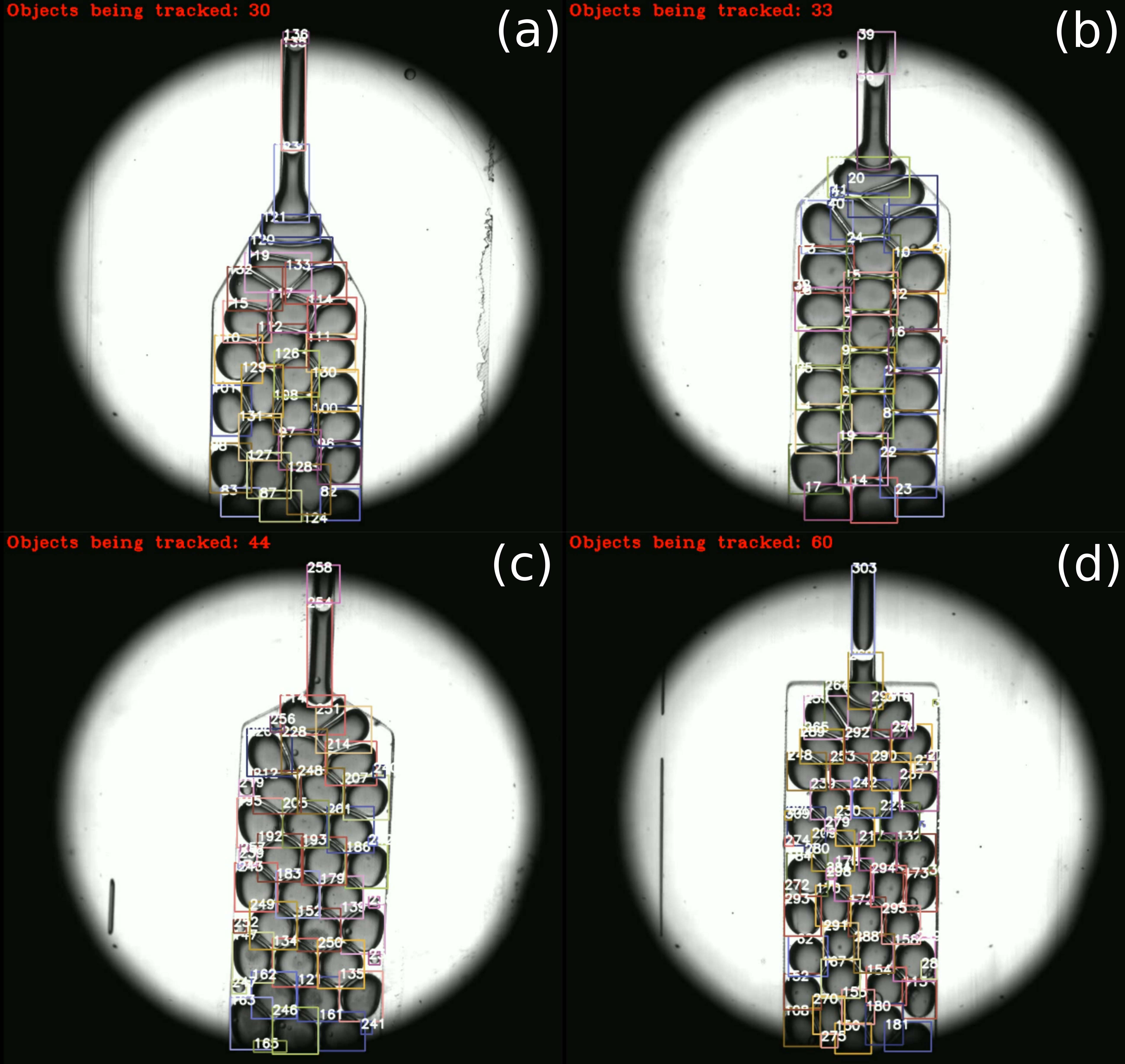}
\caption{Instances of DropTrack’s output. The water and oil injection rates are $300$ and $150 \mu$l per min respectively. The opening angle (a) $\alpha=30$, (b) $\alpha=45$ (c) $\alpha=60$ (d) $\alpha=90$ degrees. (Multimedia view). \label{fig_movie}}
\end{center}
\end{figure}

Here we shortly recap the main features of DropTrack, while we refer to the extensive literature for further details \cite{Durve1,Durve2,Durve3,Durve4}. 
The DropTrack software is based upon the amalgamation of YOLO (You Only Look Once) \cite{redmon} and DeepSORT (Deep Simple Online and Realtime Tracking) \cite{Durve1} algorithms, adeptly tailored to the specialized task of identifying and tracking droplets within image sequences. Through customization and training, DropTrack has been calibrated to precisely recognize droplets within images and extract their trajectories across consecutive frames. Consequently, the application furnishes comprehensive data, encompassing the dynamic paths of individual droplets plus the associated width and height dimensions of the bounding boxes encapsulating them. DropTrack can analyze about 30 frames per second with GPU hardware acceleration, making it suitable for in-line data analysis\cite{Durve4}. Instances of DropTrack's output is provided in Fig. \ref{fig_movie} (Multimedia view). DropTrack was employed to analyze a total of forty-four such videos depicting the flow of droplets within diverse microchannel geometries. DropTrack's output, in conjunction with temporal data, facilitate the extraction of many physical observables pertaining to the size and spatial disposition of the droplets, as we report in the next sections.

In this investigation, we systematically explore the variability in droplet volume within the context of four distinct geometric microchannels, characterized by varying opening angles $\alpha$ ranging from 30 to 90 degrees (refer to Fig.  \ref{fig_device}(a)). 
Each experiment undergoes scrutiny over a duration of 0.68 seconds, yielding a collection of 2044 frames in video format. These ones are then analysed by DropTrack, which facilitates the generation of bounding boxes encircling individual droplets, with each droplet being assigned a unique identifier. The size of each droplet is finally approximated by an ellipse encompassed by the aforementioned bounding box.

\section{Results}
\subsection{Measuring the deformability as a function of the opening angle}

\begin{figure}
\begin{center}
\includegraphics[width=\linewidth, keepaspectratio]{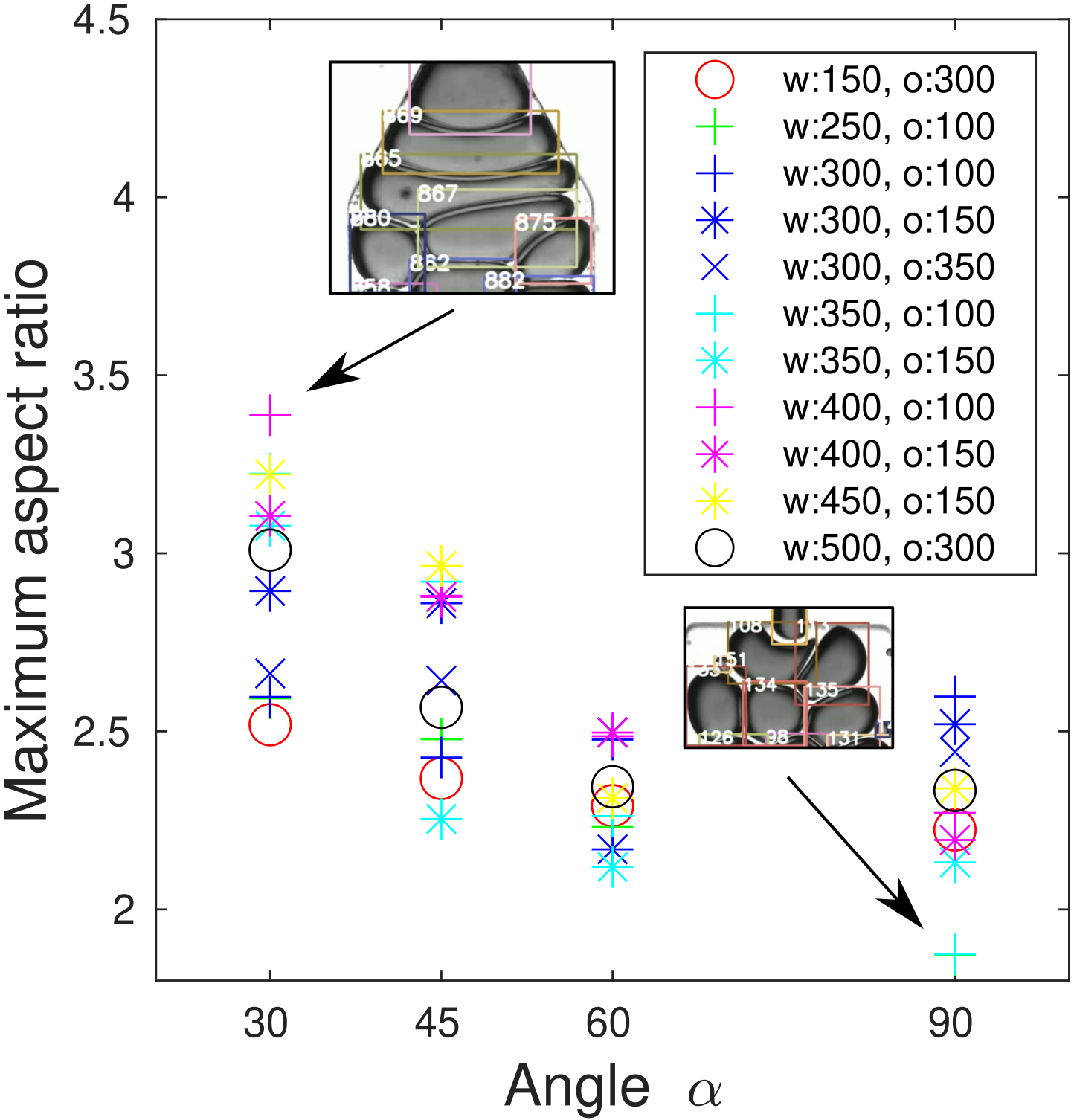}
\caption{Droplet aspect ratio for various microchannel geometries. The aspect ratio is measured as the ratio of width to height of the droplet. The inset labels report water (w) and oil (o) injection rates in $\mu l$ per min, while the images show droplets with maximal (top) and minimal (bottom) deformation for two different values of $\alpha$. Each droplet is labelled by a colored bounding box and a number. Increasing the aperture angle diminishes the deformation, basically because of a reduction of the packing density downstream. \label{fig_ar1} }
\end{center}
\end{figure}

As the liquid droplets traverse the main channel, shape deformations, mainly attributed to the presence of preceding droplets and the constrained spatial environment, start to occur, as illustrated in Fig.\ref{fig_device}a. Maximum deformation of these droplets is discerned primarily at the initial segment of the exit channel since, as they progress downstream, spatial confinement reduces and an equilibrium configuration is gradually attained. The extent of droplet deformability can be quantified in terms of the ratio of width to height within a bounding box, as identified through the DropTrack software.

In Fig.\ref{fig_ar1} we show the average maximum aspect ratio measured within each frame 
for several values of water and oil injection rates in microchannels of different opening angles $\alpha$ (as detailed in Section \ref{exp_details}). 
On a general basis, as $\alpha$ increases, there is a propensity for a decrease in deformability. This is because augmenting $\alpha$  expands the available space at the inception of the primary channel, thus reducing local packing density and, consequently, shape deformations. However, the interrelationship between these parameters follows a complex pattern. Indeed, 
the trend deviates from the observations made for opening angles exceeding $\alpha_c\simeq 60^{\circ}$ (essentially regardless of the water/oil injection rates explored in the present study) because, in these cases, the opening becomes sufficiently wide to allow the droplets to relax towards a quasi-spherical shape. 
Notably, the existence of an optimal value of $\alpha$ emerges as a key factor for the control of droplet morphology during transport, as discussed in the next section.

\subsection{Measuring the size distribution} 

\begin{figure*}
\includegraphics[width=\textwidth, keepaspectratio]{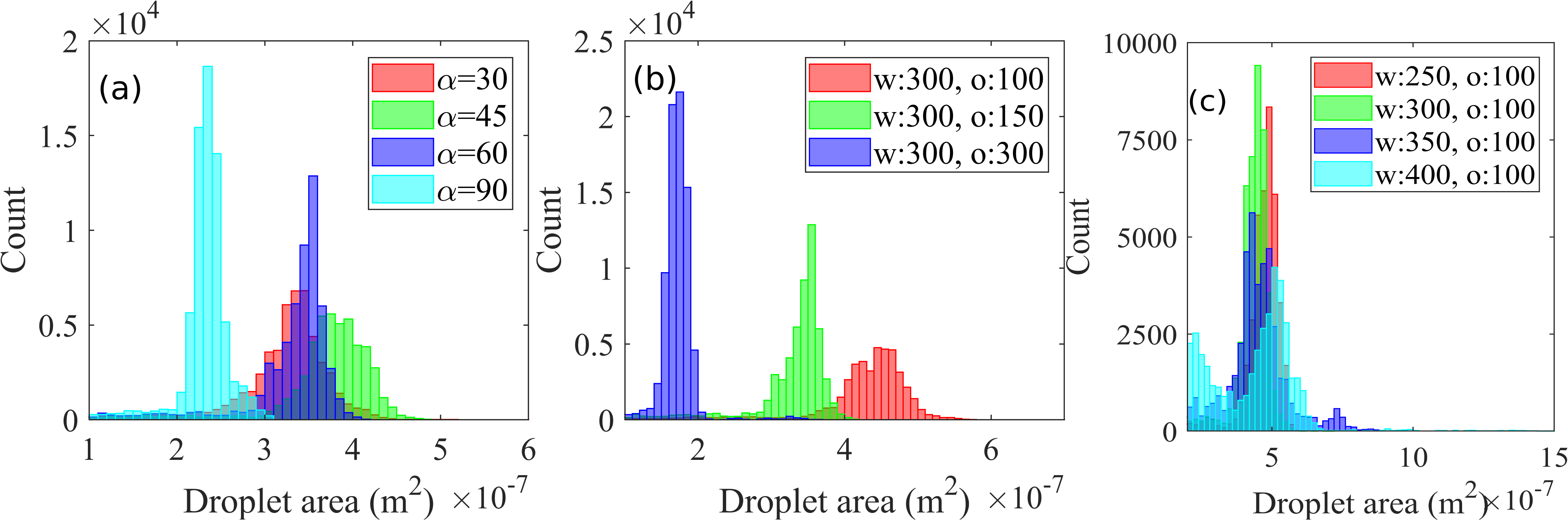}
\caption{(a) Droplet size distribution for four different values of the opening angle $\alpha$ (as shown in Fig.\ref{fig_device}a). The water injection rate is 300 $\mu l$ per min while the oil injection rate is 150 $\mu l$ per min. If $\alpha$ exceeds $60$ degrees, the droplet size undergoes a sharp decrease.   (b)-(c) Droplet size distribution for three oil injection rates (b) and four water injection rates (c). The opening angle is $\alpha=60^{\circ}$.  Increasing the oil rate diminishes droplet size; this is because the high shear at the nozzle facilitates the rupture of the jet leading to smaller droplets. On the contrary, changing the water injection rate mildly affects the size. Water (w) and oil (o) injection rates are in $\mu l$ per min. \label{fig_dd1}}
\end{figure*}

Since the experimental setup employs a camera capturing two-dimensional images, in Fig.\ref{fig_dd1}a we depict the distribution of droplet area resulting from four distinct microchannel geometries, keeping constant injection rates of oil and water across all configurations. The distribution displays a bimodal nature, delineating two discernible groups: this signifies a marked transition in droplet sizes when $\alpha$ exceeds 60 degrees.
Note that, within the subgroup constituted by opening angles $30, 45$ and $60$ degrees, the extent of the distribution spread presents a nuanced pattern with a mean around $4-5 \times$ $10^{-7}  m^2$, while for larger angles it shrinks towards lower sizes, basically because the emulsion turns almost monodisperse (with smaller circular droplets) under weaker confinement. 

Interestingly, the size distribution of the droplets is also significantly influenced by the injection rates at which oil and water are injected into the system. In Fig.\ref{fig_dd1}b-c, we show the measured droplet size distribution for various oil (b) and water (c) injection rates (while maintaining a constant rate of the other component) and for $\alpha=60^{\circ}$. Notably, the droplets exhibit a considerable reduction in size as the oil injection rate augments while an elevation of the variance as the rate decreases. This is essentially due to the increasing shear at the orifice, an effect that favours the rupture of the water jet and the formation of smaller droplets.  On the contrary, varying the water injection rate causes minor effects on droplet size (Fig.\ref{fig_dd1}c). In this case, the mean values remain relatively consistent (around $5-6\times$ $10^{-7}  m^2$) across different rates, while the distributions display an amplified degree of spread.

In summary, the results discussed so far show that, while the droplet deformation can be controlled, with good accuracy, solely by a geometrical feature of the device (i.e. the aperture angle), the size exhibits a more intricate behavior, in which both device design and injection rates of fluids play a fundamental role.

\subsection{Droplet ordering within the main channel}
\begin{figure*}
\includegraphics[width=13cm, keepaspectratio]{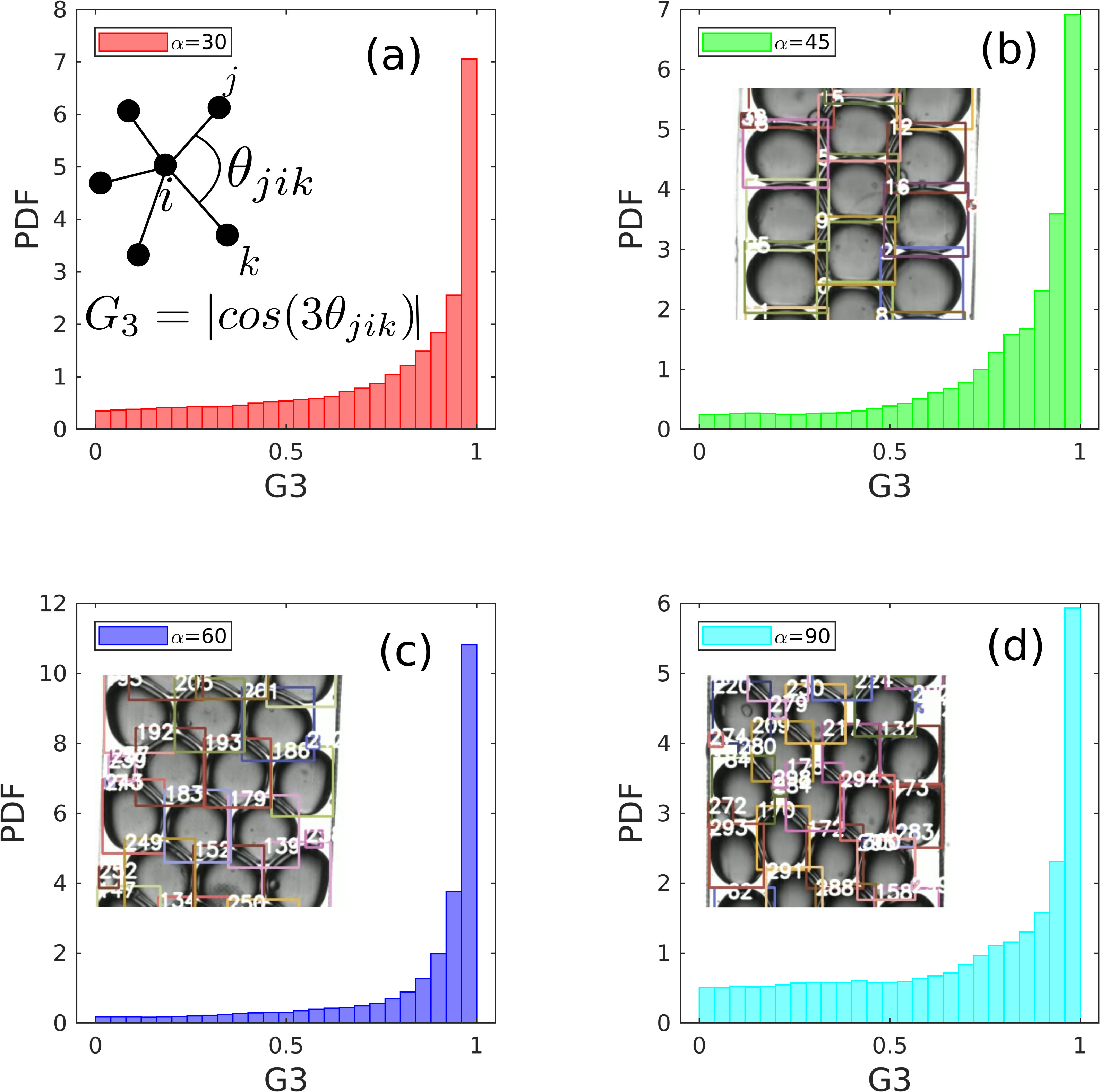}
\caption{Probability Distribution Function (PDF) of G3 order parameter for various opening angles $\alpha$ (a) 30, (b) 45, (c) 60 and (d) 90 degrees. The water injection rate is 300 $\mu l$ per min and the oil one is 150 $\mu l$ per min.  An optimal hexagonal arrangement is found for $\alpha\simeq 60^{\circ}$ where the distribution displays a pronounced peak at $1$. The fat tails observed for other values of $\alpha$ indicate a limited hexagonal order. Inset in (a): schematic illustration of the geometry of the droplets array used for the calculation of G3.
\label{fig_g1}}
\end{figure*}

Alongside discernible variations of droplet dimensions and configurations in different microchannel geometries, our results also show that, once in the channel, the droplets attain a stable ordered arrangement whose degree of organization depends, once again, on the opening angle as well as on the injection rates of the fluids. 
To achieve this, we compute the hexatic order parameter, denoted as G3, and subsequently construct its probability distribution function \cite{montessori2}. Such distribution is graphically represented with respect to the measured G3 values, defined as $G3 = \vert \cos(3 \theta_{jik}) \vert$. Here, $\theta_{jik}$ represents the angle formed by all droplet triads while maintaining the central droplet $i$ as a fixed reference point and neighboring droplets $j$ and $k$ positioned immediately adjacent to droplet $i$, as depicted in the inset of Fig.\ref{fig_g1} (a). This formulation of G3, which converges towards unity when the angle $\theta_{jik}$ assumes multiples of $\pi/3$,  serves as a robust metric for discerning whether the neighboring droplets of a given droplet $i$ conform to a hexagonal lattice arrangement.

The probability distribution function of G3 values, measured across four distinct geometries, is presented in Fig.\ref{fig_g1}. Interestingly, the microchannel characterized by an opening angle of $\alpha=60^{\circ}$ exhibits a prominent peak with G3 values reaching unity, thus suggesting that a significant proportion of the angles (as measured between droplet triads $jik$) closely approximate multiples of $\pi/3$. Also, the droplets manifest an intermediate size distribution for this specific geometry (as depicted in Figure \ref{fig_dd1}), an indication that this one represents an optimal condition for self-organization into a hexagonal lattice. In contrast, smaller droplets observed at an opening angle of $\alpha=90^{\circ}$ possess greater freedom in their movement, thereby disrupting the hexagonal order, while larger ones with an opening angle of $\alpha=45^{\circ}$ encounter spatial constraints that limit their ability to achieve hexagonal symmetry. Further detailed study is required to establish the relationship between droplet packing density and equilibrium structures in microchannels.

\begin{figure*}
\centering
\includegraphics[width=\textwidth, keepaspectratio]{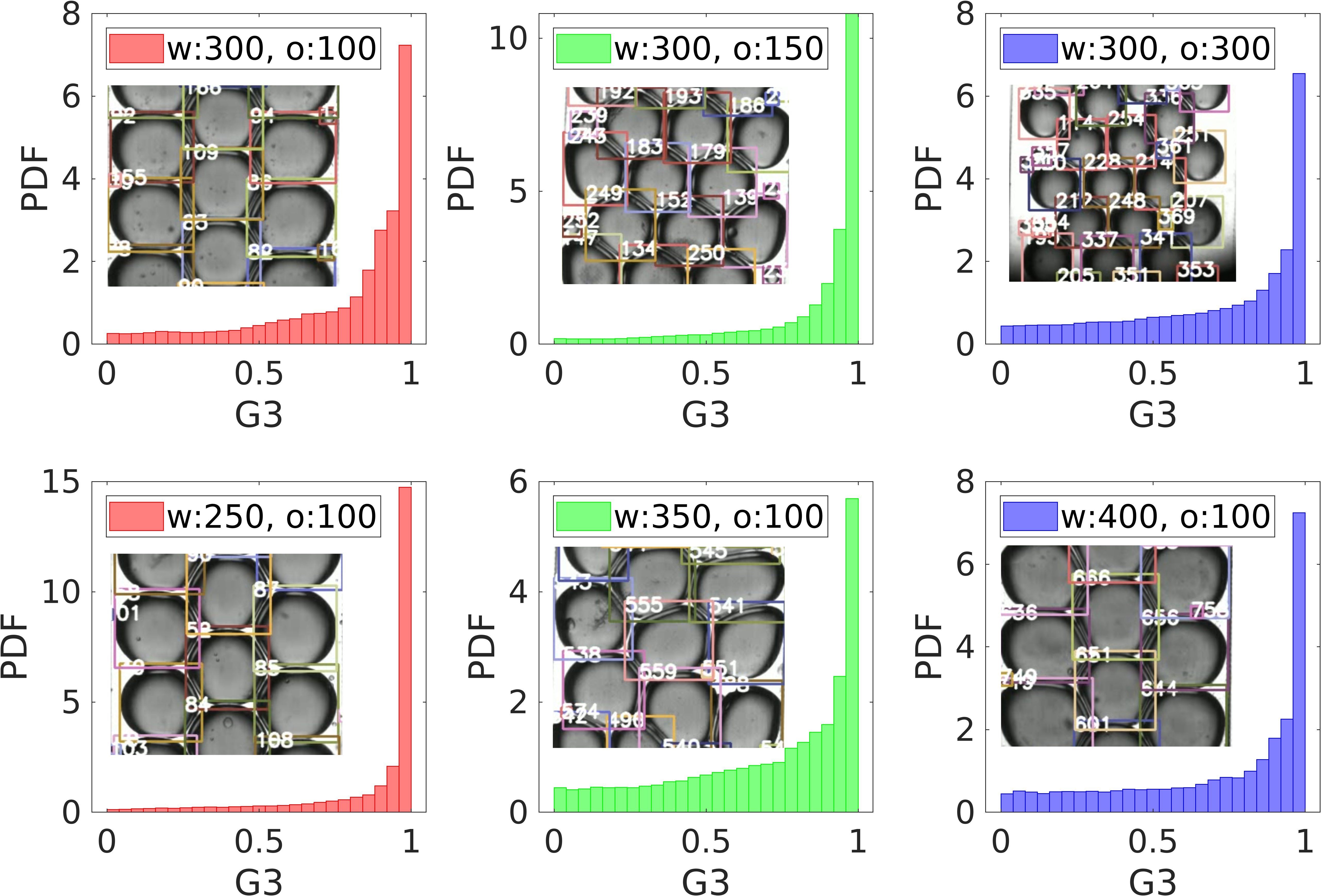}
\caption{Distribution of G3 order parameter for various water (w) and oil (o) injection rates (in $\mu l$ per min). The opening angle $\alpha$ is kept constant at 60 degrees. The insets show equilibrium droplet configuration down the main channel. 
An optimal ordered arrangement (green plot, top row and red plot, bottom row) is found for droplets of intermediate sizes displaying reduced shape fluctuations.  Illustrative images of droplets in the main channel corresponding to the water and oil injection rates are shown within the insets.\label{fig_g2}}
\end{figure*}

As previously mentioned, the propensity for hexagonal lattice formations is also affected by the water/oil injection rates. In Fig.\ref{fig_g2} we showcase the G3 distribution obtained when either oil (top row) or water (bottom row) rates are varied, for a fixed microchannel geometry. We find that the configurations characterized by intermediate droplet sizes (as depicted in the green plot of Fig.\ref{fig_dd1}b and the red one of Fig.\ref{fig_dd1}c) showcase the most conspicuous G3 order parameter peak (see the central plot, top row, and left plot, bottom row, of Fig.\ref{fig_g2}), likely because under these conditions the emulsion attains a suitable balance between droplet size and degree of confinement. Indeed, this phenomenon is particularly pronounced when the system exhibits minimal fluctuations in droplet size distributions, as illustrated for example in the red plot of Fig.\ref{fig_dd1}c and in the corresponding G3 distribution in the left lower panel of Fig.\ref{fig_g2}.

These observations collectively suggest that the emergence of a spatial order depends, in a highly non-trivial way, on an intricate interplay between droplet size, microchannel geometry, and fluid injection rates.

\subsection{Hexagonal structures and packing fraction} 
In the previous section, the calculation of the G3 order parameter takes into account all droplets in the channel, thus including the ones near the walls where perfect hexagonal symmetry is inevitably absent. Here, we assess the proportion of droplets having precisely six neighboring droplets and exhibiting a hexatic order parameter exceeding 0.98 for all triads, thus inherently excluding droplets located at the boundaries.

\begin{figure*}
\begin{center}
\includegraphics[width=\linewidth, keepaspectratio]{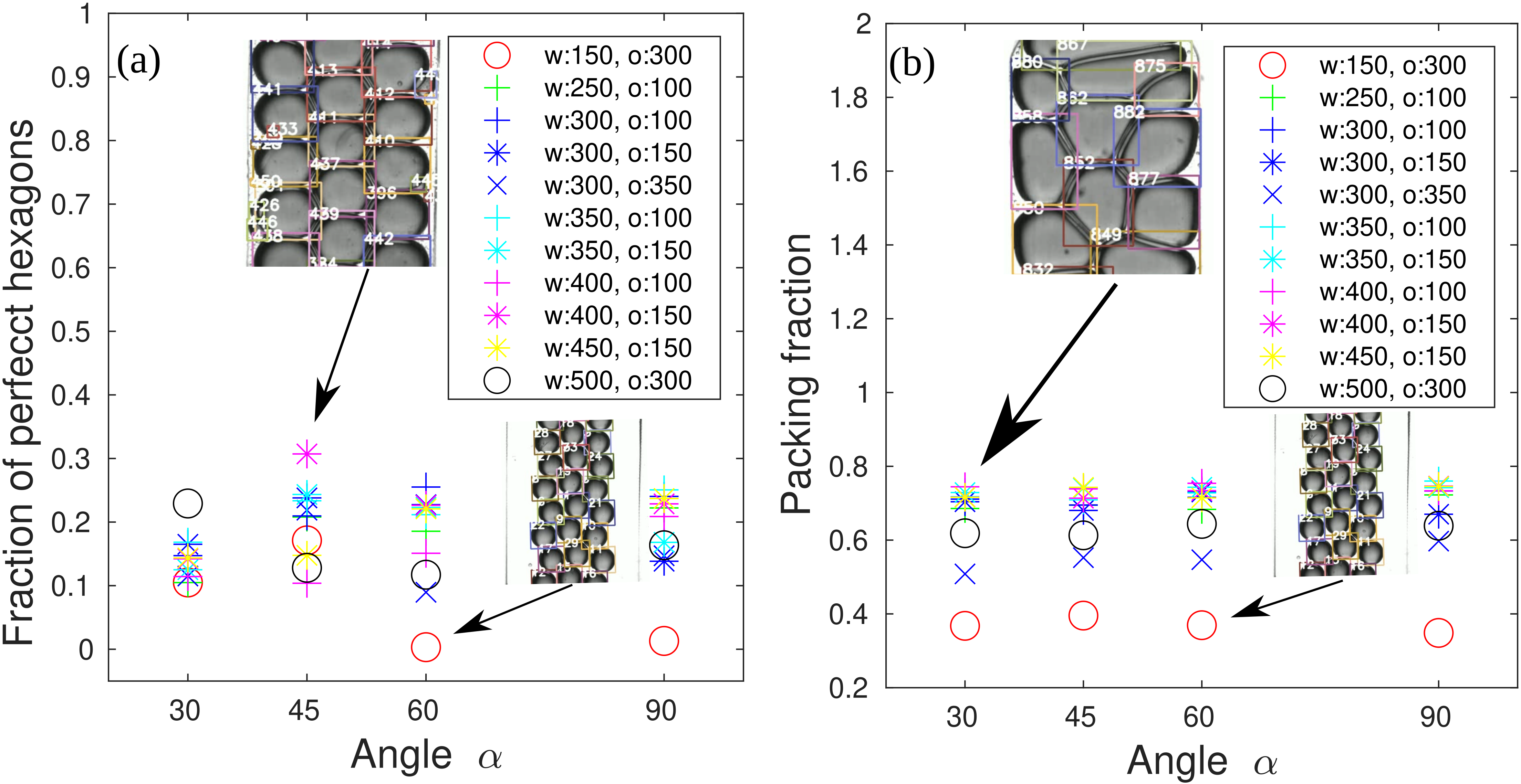}
\caption{(a) The fraction of hexagonal structures within the main channel for various combinations of experimental parameters. (b) The droplet packing fraction within the main channel for various combinations of experimental parameters. Its values are mainly controlled by the oil/water injection rates, while they are very weakly affected by the opening angle. The inset labels show the water injection rate (w) and the oil injection rate (o) in $\mu l$ per min. Illustrative images of droplets in the main channel corresponding to the maximum and minimum values of the observables are shown within the insets.\label{fig_hex1} } 
\end{center}
\end{figure*}

The variation in the fraction of droplets satisfying these criteria is visualized in Fig.\ref{fig_hex1}a across diverse experimental parameters. Unlike the previous cases, here clear and definitive trends are less discernible when considering either experiments with different opening angles or microchannel geometries featuring varying water and oil injection rates. In broad terms, the fraction of hexagonal structures in the bulk of the microchannel ranges approximately from $0.1$ to $0.35$,  an indication that, in the best scenario, only $\sim 35\%$ of droplets self-organize in a perfect hexagonal configuration. Nonetheless, a value equal to $0.35$ can be considered good enough considering the stringent criteria employed to identify these hexagonal structures (which necessitate both exactly six neighbors and a G3 higher than $0.98$) and the limited control on their formation achieved by solely changing either the injection rates of fluids or the geometrical setup. 
Finally, the higher fractions of hexagonal structures observed in larger systems (see the results at $\alpha=90^{\circ}$) is somewhat expected given the inherent nature of hexagonal packing.

A deeper insight can be gained by computing the droplet packing fraction, defined as the ratio of the cumulative area occupied by droplets (which are approximated as ellipses enclosed within identified bounding boxes) within the channel with respect to the total channel area. In Fig.\ref{fig_hex1}b we show the packing fraction in four distinct geometries and under varying flow conditions. It generally ranges from $\sim 0.3$ to $\sim 0.8$, a variation mainly due to flow rate changes rather than opening angles. More specifically, lower values are found for higher oil injection rates (see, for example, the blue circles in Fig.\ref{fig_hex1}b), basically because the large shear of the dispersed phase hinders the formation of highly packed configurations of droplets. As the oil rate turns lower, packing fractions stabilize around 0.8 for all values of $\alpha$ explored in the experiment. It is finally worthwhile to note that these packing fraction measurements are quickly estimated by DropTrack in real-time, although their precision can slightly diminish for droplets characterized by heightened deformability.

\section{Discussion}
In summary, we analyze dense emulsion microchips with varying opening angles and emulsions formed with varying oil and water injection rates. 

Our results show that, while the droplet aspect ratio is found to diminish for increasing opening angles and essentially regardless of injection rates of fluids, the size distribution of the droplets as well as their arrangement in the channel exhibit a more intricate scenario. 
For fixed values of injection rates of dispersed and continuous phase, we find a suitable aperture angle $\alpha_c$ (approximately equal to $60$ degrees in our setup) where large part of the emulsion self-organizes into a ordered structure, very likely because of an optimal balance between droplet size and degree of confinement. 
For $\alpha>\alpha_c$, droplet size turns considerably smaller and uniform, basically because shape deformations become milder under weak confinement, thus the degree of order is found to decrease. 
Changing the injection rates of the fluids (keeping constant the aperture angle) further enriches the picture. Indeed, increasing the oil injection rate (continuous phase) diminishes the droplet size due to the high shear at the orifice; on the contrary, augmenting the water injection rate (dispersed phase) negligibly affects the medium value while amplifying the spread of the size distribution. Our findings also suggest that the injection rates of both fluids can be properly tuned 
to control the formation of ordered structures in the channel. In particular, low rates of the dispersed phase favour regular hexagonal configurations with a packing fraction varying from $\sim 0.6$ to $\sim 0.8$ regardless of the confinement conditions, while higher values compromise the stability of such super-structures.

The measurement of droplet size, structural characteristics, and deformability holds paramount significance in different contexts ranging from material science for the manufacturing of scaffolds of tissues \cite{guevorkian,douezan,costantini}, to the biological domain for understanding the behavior of cell clusters crossing physiological constrictions \cite{guzowski,tiribocchi,montessori,sh_au}, up to the pharmaceutics for improving design and functioning of drug delivery products \cite{pais,pontrelli}. However, achieving precise control of droplet features (such as size distribution while adhering to defined deformability) under highly confined regimes is still a difficult task in the realm of microchip design.
In this investigation, we have elucidated the intricate interdependency among dimensions, spatial arrangement, and deformability characteristics of droplets as they flow within the primary channel of a flow-focusing microchip. This has been done using the deep-learning software DropTrack which is proven, once more, a powerful platform to understand the physics of these complex systems and to potentially predict their behavior. Its demonstrated applicability to various lab's data pertaining to different microfluidic experiments considerably limits the risk of overfitting, which occurs when models perform well within the developer's setup but poorly in others.  This is particularly important in fluid dynamics, where the complexity of the parameter space involving different scales as well as the details of the device manufacturing may crucially affect performance and results. In addition, its capability to recognize patterns and non-trivial dependencies often inaccessible to experiments is a crucial step to strengthen the model and enhance the automation of the identification/tracking procedure.

Further in-depth inquiries are imperative to address a number of questions of relevance for experiments and theory, comprising the use of ML tools in microfluidics. On the practical side, the ability to control flow rates, volume fractions and interfacial properties to augment the production of droplet-based soft materials and to standardize their manufacturing procedure remains a challenge. Much of the complexity raises from droplet deformability, which introduces non-linearities in the otherwise linear Stokes flows, together with variations of the channel geometry, which often affect the dynamics in unexpected ways \cite{anna}. Alongside these macroscopic effects, a careful assessment of the near-contact interactions governing the physics at the scales of the fluid interfaces as well as thermal properties are crucial to control the mechanical stability of the material\cite{Succi_thermal}. This highly non-trivial parameter space results in a multifarious range of flowing patterns whose features are exceptionally difficult to predict.

The integration of ML methods to computer simulations would surely contribute to tackle the complex fluid dynamics of these multiphase flows as well as to optimize design and testing, potentially leading a higher rate of experimental success and easier commercialization of microfluidic chips \cite{mctyre,laska}. However, further work is needed to build advanced ML techniques capable of quantifying, for example, the evolution of droplet shapes and break-up conditions observed under different flow rates, especially in systems composed of tightly packed droplets where morphological deformations significantly depart from elliptical-like geometries. 
Indeed, a question of high relevance concerns the requisite microchip dimensions yielding droplets of desired size and structural attributes.
These improvements could considerably extend the applicability of ML tools to other soft materials such as foam, cells and tissues. Finally, 
much efforts should be addressed to scale up the operation of ML methods to lab-on-chip platforms, in order to minimize modelling changes caused by the use of different protocols or training datasets.

\section{Conclusion}
In this work, we study the effects of opening angle and injection rates of oil (continuous phase) and water (dispersed phase) on droplet sizes and resulting emulsion structure within a fabricated microchip. The physics is captured by a number of statistical properties, such as aspect ratio, size distribution and hexatic order parameter computed using DropTrack, a ML based image analysis tool successfully adopted to identify and track droplets in microfluidic experiments. This technique, besides providing accurate and reliable results (as also demonstrated by previous works \cite{Durve1,Durve2,Durve3,Durve4}), considerably expedite the analysis, 
which would typically be a challenging task owing to the prohibitive labor-intensive nature of traditional methodologies.

The obtained results reveal that droplets undergo varied deformations prior to attaining equilibrium within distinct devices characterized by different opening angles $\alpha$. The observed diversity in deformations arises from factors such as the injection rates, available space (as determined by the opening angle $\alpha$), the droplets' sizes, the local packing fraction of adjacent droplets proximal to the main channel entrance and the characteristics of the device walls. Additionally, an optimal opening angle $\alpha$ is identified, wherein droplets organize themselves in closest proximity to a hexagonal structure for a given oil and water injection rate. This investigation encompasses the measurement of deformations, droplet size distribution, and rearrangement, elucidating the intricate relationships among these parameters concerning microchannel geometry and injection rates. The machine learning-based tool (DropTrack) used here can provide efficient and cost-effective real-time measurements of the quantities mentioned above. The findings necessitate further detailed study 
 with finer experiments and simulations for understanding these intricate dependencies, aiming to achieve better control over droplet size and rearrangements while adhering to specified maximum deformation criteria for applications in tissue engineering and other bio-transport contexts.

\begin{acknowledgments}
M.D. and S.O. contributed equally to this work. The authors acknowledge funding from the European Research Council Grant Agreement No. 739964 (COPMAT) and ERC-PoC2 grant No. 101081171 (DropTrack). M.L. acknowledges the support of the Italian National Group for Mathematical Physics (GNFM-INdAM). A.C. and D.P. acknowledge the support of the European Union by the Next Generation  EU project ECS00000017 ‘Ecosistema dell’Innovazione’ Tuscany Health Ecosystem (THE, PNRR, Spoke 4: Nanotechnologies for diagnosis and therapy). We gratefully acknowledge the HPC infrastructure and the Support Team at Fondazione Istituto Italiano di Tecnologia.
\end{acknowledgments}

\section*{Data Availability Statement}
The data that support the findings of this study are available from the corresponding author upon reasonable request.

\section{Conflict of Interest Statement}
The authors have no conflicts to disclose.

\section{CRediT (Contribution Role Taxonomy) Statement}
\noindent Conceptualization – M.D., M.L., S.O., A.C., J-M. T., A.M., S.S. (equal) \\
Data Curation – M.D., S.O., A.C. (equal) \\
Formal Analysis – M.D., A.T., M.L. (equal) \\
Funding Acquisition – A.C., D.P., M.L., S.S. \\
Methodology – M.D., S.O., A.C., A.T., M.L. (equal) \\
Project Administration – S.S. \\
Software – M.D., M.L., A.T., A.M. (equal) \\
Supervision – S.S., A.C., D.P. \\
Visualization – M.D., A.T., M.L. \\
Writing/Original Draft Preparation – A.T., M.D., J-M. T., A.C., S.O. \\
Writing/Review \& Editing – All authors (equal)

\bibliography{biblio}

\end{document}